\newcommand{\beq}{\begin{equation}}
\newcommand{\eeq}{\end{equation}}
\newcommand{\eq}[1]{\begin{align}#1\end{align}}
\newcommand{\lf}{\left}
\newcommand{\rf}{\right}
\newcommand{\nt}{\notag}

\documentclass[aps,prb,twocolumn,aps,superscriptaddress]{revtex4-1}
\usepackage{graphicx}
\usepackage{verbatim}
\usepackage{mathrsfs}
\usepackage{cancel}
\usepackage{bm}
\usepackage{float}
\usepackage{color}
\usepackage{amsmath,amsfonts,amssymb}
\usepackage{hyperref}
\usepackage{cleveref}
\usepackage{epstopdf}

\begin{document}

\title{Spin Hall conductivity in topological Dirac semimetals}

\author{
Katsuhsia Taguchi$^1$, Daisuke Oshima$^2$, Yusuke Yamaguchi$^2$, Tatsuki Hashimoto$^1$, Yukio Tanaka$^2$, Masatoshi Sato
}

\affiliation{
Yukawa Institute for Theoretical Physics, Kyoto University, Kyoto 606-8502, Japan
\\
$^2$ Department of Applied Physics, Nagoya University, Nagoya 464-8603, Japan 
}

%
%
%
%
%

\date{\today}
%

\begin{abstract}
We theoretically investigate the spin Hall conductivity (SHC) in topological Dirac semimetals (TDSMs) whose Dirac points are protected by rotational symmetry.
On the basis of a general phase diagram of the system with time-reversal, inversion and four-fold rotational symmetries, 
we reveal that the SHC is sensitive to the phase to which the system belong. 
The phase and the SHC are characterized by the mirror Chern numbers and the presence or absence of gapless bulk Dirac points.
It is also found that the representative TDSM Cd$_3$As$_2$ supports a large and negative SHC $\sigma_{xy}^z\sim -10^4 (\hbar/e) (\Omega.\textrm{m})^{-1}$.
The principle behind the dependency of SHC on the phase diagram is also explained. 
\end{abstract}

\maketitle

\section{Introduction}
Topological Dirac semimetals (TDSMs) are 3D materials having both a linear dispersion along all momentum directions and a nonzero topological invariant. 
Their Dirac points (DPs)(gapless points) are protected by rotational symmetry, and they exist stably along the rotational symmetric axis in a finite parameter region unlike an accidental Dirac semimetal existing at 
the topological phase transition point between a 3D topological insulator and a normal insulator\cite{Yang14}.
TDSMs can be distinguished from ordinary Dirac semimetals (DSMs) by the mirror Chern numbers $n_{\mathcal{M}}$ at $k_z=0$ and $\pi$ plane; 
$n_{\mathcal{M}}=0$ in DSMs but $n_{\mathcal{M}}\neq0$ in TDSMs.
To date, TDSMs with rotational symmetry have been synthesized and experimentally demonstrated\cite{Liu2014a, Liang2014,He2014,Ali2014,Uchida2017}.

Several characteristic phenomena specific to TDSMs have been studied theoretically and experimentally\cite{Liu2014a, Liang2014,He2014,Ali2014,Uchida2017, Burkov2016a,Araki2018}.
Among them, the spin Hall effect, {\it i.e.} spin current generation via an applied electric field, is intriguing. 
The spin Hall effect is a central issue in spintronics, and the non-dissipative spin current has a possible application to energy-saving electronic devices \cite{Hoffmann2013,Sinova2015}.

Whereas the spin Hall conductivity (SHC) of TDSMs has been discussed in the low-energy model Hamiltonian, \cite{Burkov2016a} a full lattice description is necessary to capture the whole phase diagram of TDSMs: \cite{Yang14}
A general 3D lattice Hamiltonian 
possesses multiple phases, such as TDSM, DSM, and weak topological (crystalline) insulators \cite{Yang14},
which are not reproduced in the low-energy model. 
Moreover, the low-energy description is known to have a subtle ambiguity in transport phenomena \cite{Vazifeh2013,Goswami2013}.

In this paper, we evaluate the SHC systematically based on a lattice model with time-reversal, inversion and four-fold rotational symmetries. 
First, we consider the low-energy limit of the model Hamiltonian. 
We derive the SHC analytically, and find that  
the SHC is proportional to the distance between DPs along the rotational symmetric axis ($\Gamma$-$Z$ axis).
The low-energy Hamiltonian accidentally preserves the spin in the $z$-direction, from  which the simple relation is obtained. 
Then, we evaluate the SHC by using the full lattice model. 
We reveal that the SHC is sensitive to the phase to which the system belong. 
The phase and the SHC are characterized by the mirror Chern numbers and the presence or absence of gapless bulk Dirac points.
We also find that the SHC obtained in the low-energy limit reproduces that in the lattice model only in a part of the TDSM phase. 
The principle behind the dependency of SHC on the phase diagram is explained. 
Furthermore, since the SHC is given by the quantized unit of electrical conductance and the distance between DPs, we estimate that the representative TDSM Cd$_3$As$_2$ supports a large and negative SHC $\sigma_{xy}^z\sim -10^4 (\hbar/e) (\Omega.\textrm{m})^{-1}$.

This paper is organized as follows. 
In Sec.\ref{sec:Model}, we introduce a general model Hamiltonian of TDSMs that supports time-reversal, inversion and four-fold rotational symmetries.
We also review the phase diagram of the system. 
In Sec.\ref{sec:III}, we calculate the SHC using the Green's functions techniques in the clean limit. 
In Sec.\ref{sec:lowenergy}, we derive the analytical formula for the SHC in the low-energy description near the $\Gamma$ point. 
The mixing effects in the low-energy description is discussed in Sec.\ref{sec:mixing}.
We provide the result of the SHC for the full lattice Hamiltonian in Sec.\ref{sec:lattice}.
In Sec. \ref{sec:IV}, we compare the full lattice calculation with the low-energy description. 
We also identify the topological invariant that explains a jump of the SHC at  $M/t_z=1$.
Finally, we estimate the SHC for a candidate material Cd$_3$As$_2$.
%

\section{Model}\label{sec:Model}
To calculate the SHC, we consider a general Hamiltonian for TDSMs having time-reversal, inversion, and four-fold rotational symmetries.
Assuming that the Hamiltonian consists of orbitals with opposite parity under inversion, 
we obtain the minimal Hamiltonian as follows,\cite{Yang14,Kobayashi15,Hashimoto2016}  
\eq{\nt 
H (\bm{k}) =& a_1(\bm{k}) \sigma^x s^z + a_2(\bm{k}) \sigma^y s^0 + a_3(\bm{k})\sigma^z s^0
	\\ \label{Model-eq:1} 
	& + a_4(\bm{k}) \sigma^x s^x 
	+ a_5(\bm{k}) \sigma^x s^y, 
}
where $a_{i=1,2,3,4,5}$ is a real function given by 
\eq{ 	\label{Model-eq:2} 
	 a_{1}(\bm{k}) & = \eta  \sin{k_x}
			 \\ \label{Model-eq:3} 
	 a_{2}(\bm{k})  & = - \eta \sin{k_y}   
	  \\ \label{Model-eq:4} 
	 a_3 (\bm{k}) & = M - t_{xy} (\cos{k_x} + \cos{k_y}) - t_{z} \cos{k_z}.
	\\ \label{Model-eq:5} 
	a_{4} (\bm{k}) & = (\beta +\gamma)  \sin{k_z} (\cos{k_y} - \cos{k_x}) 
			\\ \label{Model-eq:6} 
	a_{5} (\bm{k}) 	&=  - (\beta -\gamma)  \sin{k_z} \sin{k_y} \sin{k_x}.
}
Here, $\eta$ indicates the nearest neighbor hopping amplitudes in the $xy$ plane;
$\beta +\gamma$ represents the next nearest neighbor hopping amplitudes in the $yz$ and $zx$ planes;
$\beta -\gamma$ indicates the hopping process along the body-diagonal direction of the cubic lattice; 
$\sigma^{i=x,y,z}$ and $s^{i=x,y,z}$ are Pauli matrices in the orbital and the spin spaces, respectively;
$M$ indicates the on-site potential difference between the orbitals; 
$t_{xy}$ ($t_{z}$) denotes the difference in hopping amplitudes between different orbitals along the $x$ and $y$ directions (along $z$ direction).
The Hamiltonian describes the TDSM phase in Cd$_3$As$_2$ where the orbital and the spin are given by  $|s,\uparrow, J_z=1/2 \rangle$, $|p_x+ip_y,\uparrow,J_z=3/2 \rangle$, $|s,\downarrow,J_z=-1/2 \rangle$, and $|p_x-ip_y,\downarrow, J_z=-3/2 \rangle$ 
($\uparrow$ and $\downarrow$ represent the spin degrees of freedom).

The Hamiltonian hosts time-reversal, inversion, and four-fold rotational symmetries
\begin{align}
&\mathcal{T} H(\bm{k}) \mathcal{T}^\dagger = H(-\bm{k}),\\    
&\mathcal{P} H(\bm{k}) \mathcal{P}^\dagger = H(-\bm{k}),\\
&\mathcal{C}_4 H(k_x, k_y, k_z) \mathcal{C}^\dagger_{4} = H(k_y, -k_x, k_z),
\end{align}
with $\mathcal{T}=i \sigma^0 s^y \mathcal{K}$, $\mathcal{P}=\sigma^z s^0$,
and $\mathcal{C}_4 =e^{i(2\sigma^0-\sigma^z)s^z\pi/4}$, respectively.
Combining inversion and  rotation symmetries, we also have mirror reflection symmetry 
\begin{align}
\mathcal{M}_{xy} H(k_x, k_y, k_z) \mathcal{M}^\dagger_{xy} = H(k_x, k_y, -k_z)    
\end{align}
with $\mathcal{M}_{xy} = \mathcal{P}\mathcal{C}_4^2=i \sigma^0 s^z$.
The four-fold rotational symmetry prohibits the mixing between different orbitals at a rotation symmetric line, which makes it possible to obtain stable DPs. 
Because of the mirror reflection symmetry, 
we can introduce the mirror Chern numbers at $k_z=0$ and $k_z=\pi$, respectively. 
Note that the mirror Chern numbers are well-defined even if DPs exist, as long as DPs are not on a mirror invariant plane.

\begin{figure}[tb]
\includegraphics[width=8.4cm]{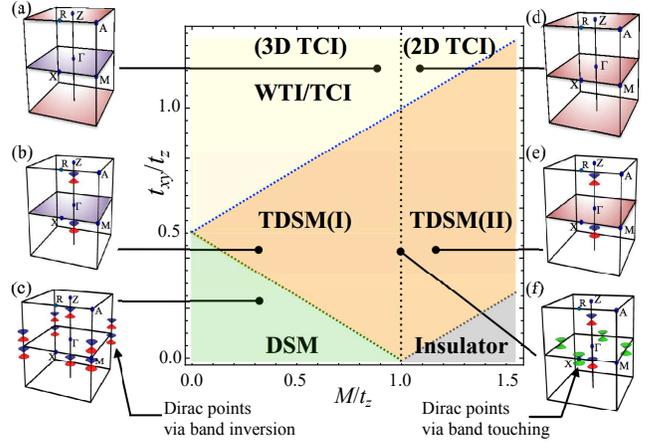}
\caption{ (Color online)  Phase diagram of the system, where $M$ and $t_{xy}$ ($t_{z}$) denotes the on-site potential difference between the orbitals and the difference in hopping amplitudes between different orbitals along the $x$ and $y$ directions (along $z$ direction), respectively\cite{Yang14}.
The phases are characterized by the mirror Chern numbers and the presence or absence of DPs. 
(a)-(f)indicate the location of DPs via band inversion (represented by red and blue cones) and band touching (represented by green cones) in the Brillouin zone.
Here, the band-touching DP are on the boundaries of the phases (dotted lines in the phases). 
At $\beta+\gamma=0$, DPs exist even at the $X$-point in the $M/t_z<1$ regime. 
The red (blue) plane in the Brillouin zone shows the nonzero mirror Chern number $n_{\mathcal{M}} =+1 (-1)$ at in $k_z=0$ and $\pi$ plane. 
}
\label{fig:1} 
\end{figure} 

Following Ref.\onlinecite{Yang14}, 
we present the phase diagram of the model Hamiltonian in Fig.\ref{fig:1}.
Each phase is characterized by the mirror Chern numbers and the presence or absence of DPs.
For $M/t_z+2t_{xy}/t_z-1<0$ (DSM phase), DPs  both at $\Gamma$-$Z$ and $M$-$A$ lines in the Brillouin zone [Fig. \ref{fig:1}(c)], and 
the mirror Chern numbers are zero both at $k_z=0$ and $\pi$. 
For $M/t_z \pm 2t_{xy}/t_z \mp 1 >0$ and $M/t_z - 2 t_{xy}/t_z - 1<0$ (TDSM phase),  DPs are located only at a line between $\Gamma$ and $Z$ points, and the mirror Chern number is nonzero at $k_z=0$ [Fig.~\ref{fig:1}(b),(e)]. 
Finally, in $M/t_z-2t_{xy}/t_z+1<0$ (weak topological insulator (WTI)/topological crystalline insulator (TCI) phase), the system is fully gaped in the bulk and has nonzero mirror Chern numbers [Fig.~\ref{fig:1}(a),(d)]. 

We notice that
the mirror Chern number at $k_z=0$ changes by a factor of two at $M/t_z=1$ line in the TDSM and WTI/TCI phases. 
The change is caused by a band touched Dirac dispersion at the $X$ point for $M/t_z=1$ [Fig. \ref{fig:1}(f)].
The difference in the mirror Chern number refines both the TDSM and WTI/TCI phases.
In particular, the WTI/TCI phase is refined into a 3D TCI [Fig.~\ref{fig:1}(a)] and a 2D TCI [Fig.~\ref{fig:1}(d)], respectively.
We denote the refined phases as TDSM(I)[Fig.~\ref{fig:1}(b)], TDSM(II)[Fig.~\ref{fig:1}(e)], 3D TCI[Fig.~\ref{fig:1}(a)], 
and 2D TCI[Fig.~\ref{fig:1}(d)], respectively. 



In addition to DPs protected by rotation symmetry, the system may have accidental DPs. 
Such accidental DPs appear along the $X$-$R$ line when $M/t_z<1$ and $\beta+\gamma=0$.
On the $X$-$R$ line, the energy of the system is given by $E(0,\pi,k_z) = \pm \sqrt{a_3^2(k_z) +a_4^2(k_z)}$, which is gapless when $k_z=\pm \arccos{(M/t_z)}$ and $\beta+\gamma=0$.
For a non-zero $\beta+\gamma$, a gap opens at the gapless point, but it stays narrow for small $\beta+\gamma$.
As we show later, such a narrow gap may provide a substantial contribution to the SHC.

\section{Spin Hall conductivity}\label{sec:III} 
In this section, we show the SHC of the model Hamiltonian $H(\bm{k})$. 
First, using the standard technique of the Keldysh Green's functions, we provide that the SHC is represented by the momentum integral of the spin Berry curvature and the Fermi distribution functions [Eq. (\ref{Model-eq:10})]. 
Second, we calculate the SHC using Eqs. (\ref{Model-eq:10})-(\ref{Model-eq:14}) and two different descriptions of the model;
the low-energy description near DPs, and the full lattice description. 
The SHC is obtained analytically in the low-energy description near DPs, and we find that the SHC is proportional to the distance between DPs if the mixing between DPs is taken into account.
We also perform the numerical calculation of the SHC in the lattice description of the model, and evaluate the phase dependence of the SHC.

\subsection{Spin Hall conductivity} \label{sec:3-1} 
Spin current density is defined by\cite{Sinova2015}
\eq{ \label{SHC-eq:1} 
j_{\textrm{s},i}^\alpha & = \langle \psi^\dagger v_{\textrm{s},i}^\alpha \psi  \rangle, 
}
where $v_{\textrm{s},i}^\alpha = \{v_i, \frac{\hbar}{2} s^\alpha \sigma^0  \}/2$ is the velocity operator of the spin current, 
$v_i = \partial H/(\hbar\partial k_i)$ is the velocity operator, and 
$\psi^\dagger = (\psi_{s,\uparrow}^\dagger, \psi_{p_x+ip_y,\uparrow}^\dagger, \psi_{s,\downarrow}^\dagger, \psi_{p_x-ip_y,\downarrow}^\dagger)$ denotes the creation operator for the model Hamiltonian $H (\bm{k})$ on the spin degrees of freedom and the orbital degrees of freedom.
The superscript and the subscript of the spin current $j_{\textrm{s},i=x,y,z}^\alpha$ indicate the polarization of spin and the direction of its flow, respectively.

The spin current in a linear response caused by an external electric field along the $y$ direction ($E_{\textrm{ex},y}$) is given by 
\eq{ \label{SHC-eq:2} 
\sigma_{xy}^z 
	& \equiv j_x^z / E_{\textrm{ex},y}. 
} 
The SHC can be represented by the standard technique of Green's functions as follows:\cite{Tanaka2008,Sinova2015}
\eq{ \label{SHC-eq:3} 
\sigma_{xy}^z 
	& = \frac{e\hbar}{\Omega} \sum_{\bm{k},\omega} 
	\textrm{tr} \lf [ v_{\textrm{s},x}^z G(\bm{k},\omega)  v_y G(\bm{k},\omega+\Omega)  \rf]^<_{\Omega\to0}, 
} 
where $e$ represents the elementary charge of electrons, and $\Omega (\to 0)$ represents the frequency of the applied electric field $E_{\textrm{ex},y}$. 
Here, $G(\bm{k},\omega)$ denotes the Green's functions of $H (\bm{k})$, and the superscript ``$<$" indicates the lesser component of the Green's function. 
The SHC $\sigma_{xy}^z $ can be decomposed into two terms, i.e., $\sigma_{xy}^z =\sigma_{xy}^{z, \textrm{ra}} + \sigma_{xy}^{z, \textrm{aa}}$, 
where $\sigma_{xy}^{z, \textrm{ra}}$ and $\sigma_{xy}^{z, \textrm{aa}}$ are constructed by the products of the retarded and the advanced Green's functions, and by only the advanced Green's functions, respectively, as follows:
\eq{ \label{Model-eq:8} 
\sigma_{xy}^{z, \textrm{ra}} & = e\hbar \sum_{\bm{k},\omega} \frac{\partial f (\omega) }{\partial \omega} 
	\textrm{tr} \lf [ v_{\textrm{s},x}^z G^\textrm{r}(\bm{k},\omega)  v_y G^\textrm{a}(\bm{k},\omega) \rf], 
	\\ \nt 
\sigma_{xy}^{z, \textrm{aa}} & = \frac{e\hbar}{2} \sum_{\bm{k},\omega}  f (\omega) 
	 \textrm{tr} \biggl[  v_{\textrm{s},x}^z \biggl( \frac{\partial G^\textrm{a}(\bm{k},\omega)  }{\partial \omega}  v_y G^\textrm{a}(\bm{k},\omega)  
	 	\\ \label{Model-eq:9} 
		& \ \ \ \ \ \ \ \ \ \ \ \ \  -   G^\textrm{a}(\bm{k},\omega)   v_y \frac{\partial G^\textrm{a}(\bm{k},\omega) }{\partial \omega}  \biggr)  - \textrm{h.c.} \biggr],
}
where $f$ is the Fermi distribution function, $G^\textrm{r}(\bm{k},\omega) = [\hbar\omega - \mathcal{H}+i\epsilon]^{-1}$ and $G^\textrm{a}(\bm{k},\omega) = [G^\textrm{r}(\bm{k},\omega)]^\dagger$ denote the retarded and the advanced Green's functions, respectively.  
These SHCs are analytically calculated in the clean limit ($i.e., $ $\epsilon\to0_+$) as shown in the Appendix \ref{sec:Appendix-A}. 
Then, it becomes 
\eq{  \label{Model-eq:10} 
\sigma_{xy}^{z} &  = e \hbar \sum_{\bm{k}} \lf[ f(E) - f(-E) \rf] \Omega_{xy}^z, 
}
where $f(E) = [1+\exp{[(E-\mu)/(k_B T)]}]^{-1}$ denotes the Fermi distribution function and $\Omega_{xy}^z$ represents the spin Berry curvature; $E$, $\mu$, $k_B$, and $T$ represent the eigenvalue, chemical potential, Boltzmann's constant, and temperature, respectively. 
The spin Berry curvature $\Omega_{xy}^z $ can be obtained by calculating $\textrm{Tr}[...]$ in Eqs. (\ref{Model-eq:8}) and (\ref{Model-eq:9}). 
The spin Berry curvature $\Omega_{xy}^z$ is given by  
\eq{\label{Model-eq:13} 
\Omega_{xy}^z    
	=&  \sum_{E_n> E_{m\neq n}} i \frac{ \langle n | v_{\textrm{s},x}^z  | m \rangle \langle m | v_y  | n  \rangle - (x \leftrightarrow y) }{ [ E_n(\bm{k}) - E_m(\bm{k})]^2  }
	\\ \nt 
	= &  \frac{\eta^2 }{2\hbar E^3} 
		\biggl[
		(M - t_{z} \cos{k_z}) \cos{k_x} \cos{k_y} 
		\\ \label{Model-eq:14} 
	& \ \ \ \ \ \  \ \  \ \ \ \ \ \  \ \ \ 	- t_{xy} (\cos{k_x} + \cos{k_y})
	      \biggr], 
}
where $ | n  \rangle$ denotes the wave function for $E_n (=E = \sqrt{a_1^2+a_2^2+a_3^2+a_4^2+a_5^2})$ of the $n$-th band for the Hamiltonian $H(\bm{k})$ in Eq. (\ref{Model-eq:1}).
It must be noted that, at $\beta=\gamma=0$, the $4\times4$ model Hamiltonian can be decomposed into a $2\times2$ block diagonal in the spin $\uparrow$ and $\downarrow$ sectors, and the Berry curvatures for the $\uparrow$ and $\downarrow$ sectors $\Omega_{xy}^\uparrow$ and $\Omega_{xy}^\downarrow$ are defined, respectively.  
Then, we found that $\Omega_{xy}^z = \Omega_{xy}^\uparrow - \Omega_{xy}^\downarrow$, because the spin is conserved at $\beta=\gamma=0$.

\begin{figure*}[t] 
\includegraphics[width=15.6cm]{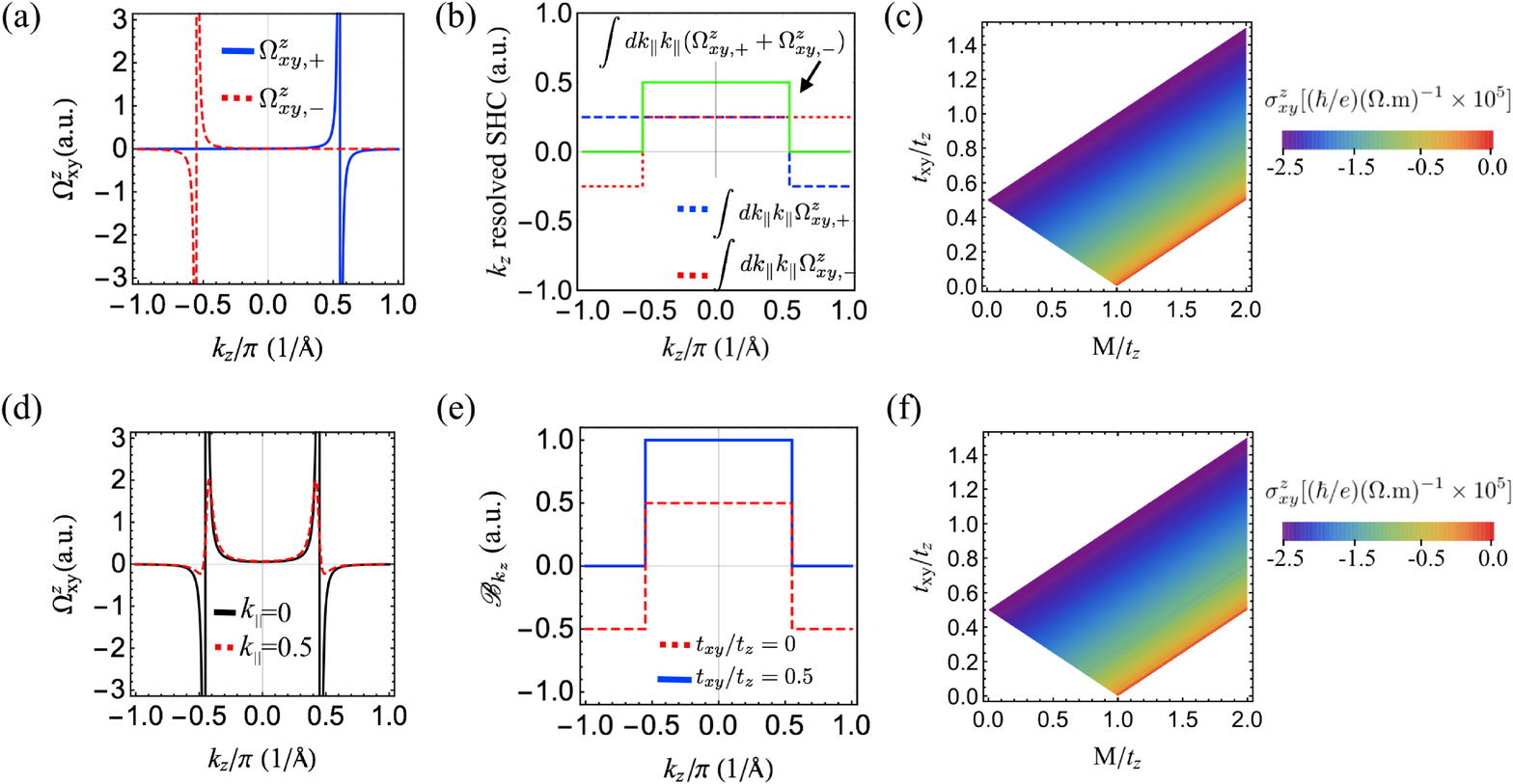}
\caption{ (Color online) 
Spin Berry curvature and SHC in the low-energy description: 
(a) Momentum dependence of the spin Berry curvature near DPs, (b) $k_z$ resolved SHC [$\int dk_\parallel k_\parallel \hbar( \Omega_{xy,+}^z + \Omega_{xy,-}^z )$], and (c) SHC in the TDSM phases in the low-energy description near the DPs, with $k_\parallel=\sqrt{k_x^2+k_y^2}$.
(d) Momentum dependence of the spin Berry curvature near the $\Gamma$ point, (e) $k_z$ resolved SHC $\mathscr{B}_{k_z}$, and (f) SHC in TDSM phases in the low-energy description near the $\Gamma$ point.  
It is noticed that, at the $t_{xy}\to0 (m_2\to0)$ limit in (f), SHC becomes infinity. 
This result implies that the low-energy description does not work when $m_2=0$.
Here, we used the parameters $\eta=0.89$ eV, $t_z/\eta =-3.4$, and $t_{xy}/t_z=M/t_z=0.5$, $a\approx$ 1\AA, and the conductance quantum $2e^2/h=7.75\times 10^{-5} \Omega^{-1}$.
The corresponding plot at that point ($t_{xy}\to0$) was excluded because the magnitude of SHC diverges at $t_{xy}$ = 0 ($m_2=0$).
}
\label{fig:2} 
\end{figure*} 

\subsection{Low-energy description near Dirac points}
\label{sec:lowenergy}
Next, we analytically obtain the SHC in the low-energy effective Hamiltonian near DPs. 
Here we assume that DPs are located near the $\Gamma$ point. 
The Hamiltonian is given by replacing $\sin{k_i} \to k_i +o(k_i^3)$ and $\cos{k_i} \to 1- \frac{1}{2}k_i^2+o(k_i^3)$ in Eq. (\ref{Model-eq:1}) as follows\cite{Kobayashi15,Hashimoto2016,Burkov2016a}: 
\eq{ \label{SHC-Ceq:1} 
\mathcal{H}(\bm{k}) & = \hbar v (k_x s^z\sigma^x  - k_y s^0\sigma^y ) + m(\bm{k}) s^0\sigma^z, 
}
where $v = \eta a/\hbar $ and $a$ represent the velocity and the lattice constant, respectively. 
Here, $m(\bm{k})$ is a parameter that represents the band inversion and is given by\cite{Burkov2016a} 
\eq{ \label{SHC-Ceq:2} 
m(\bm{k})	& = -m_0   + m_1 k_{z}^2+ m_2 (k_{x}^2 + k_{y}^2) 
}
with $m_0 = 2t_{xy} + t_z -M $, $m_1= t_{z}a^2/2 $, and $m_2= t_{xy}a^2/2 $. 
Here $m_1 m_0 >0$ is required to obtain DPs.
Below, we simply set the lattice constant $a=1$. 

Using Eqs. (\ref{SHC-Ceq:1})-(\ref{SHC-Ceq:2}), we find that DPs are located at $\bm{k}_\pm =\pm (0,0,k_0)$ with $k_0\equiv \sqrt{m_0/m_1}$. 
Then, the Hamiltonian of Eq. (\ref{SHC-Ceq:1}) is linearized by expanding near the DPs $\bm{k}_\pm$, and the linearized Hamiltonian of each DP $H_\pm $ is given by 
\eq{
H_{\pm} & = \hbar v (k_x s^z\sigma^x  - k_y s^0\sigma^y) \pm \hbar \tilde{v} (k_z \mp k_0) s^0\sigma^z 
\label{eq:linear DH}
}
with $\tilde{v} = 2m_1 k_0/\hbar = \textrm{sgn}(m_1)2\sqrt{m_0m_1}/\hbar$.
Using the Hamiltonian $H_{\pm}$, the SHC is given by, 
\eq{
\sigma_{xy}^{z}  & = \sigma_{xy,+}^{z}   +\sigma_{xy,-}^{z}, 
}
where $\sigma_{xy,\pm}^{z}$ is the SHC for the Hamiltonian $H_\pm$.
This is a linearized model Hamiltonian near the DPs along the $\Gamma$-$Z$ direction. 
Here, $\sigma_{xy,\pm}^{z}$ can be individually calculated based on the assumption of no interband interaction (no mixing effect) between each $H_\pm$.
This SHC $\sigma_{xy,\pm}^{z}$ in the clean limit is obtained by a similar calculation as that in Eqs.  (\ref{Model-eq:10})-(\ref{Model-eq:13}) as follows:
\eq{
\sigma_{xy,\pm}^{z} &= e \hbar \sum_{\bm{k}}  [f(E_\pm) - f(-E_\pm) ] \Omega_{xy,\pm}^z , 
\label{eq: sum}
} 
where $E_\pm = \hbar \sqrt{v^2 (k_x^2+k_y^2) + \tilde{v}^2 (k_z \mp k_0)^2 }$ and 
$\Omega_{xy,\pm}^z = \pm \hbar v^2 \tilde{v} (k_z \mp k_0)/(2E^3_\pm) $ [as shown in Fig.\ref{fig:2}(a)] denote the eigenvalue and the spin Berry curvature for $H_\pm$, respectively.  
We perform the summation in Eq.(\ref{eq: sum}) in a rotational symmetric manner along the $k_z$-axis.
Because the system is a 2D spin Hall insulator for a fixed value of $k_z$ (except for $k_z =\pm k_0$), the 2D SHC is quantized to $ (e^2/h) \times \mathbb{Z}$. As a result, the SHC in the overall system is given as follows (Appendix \ref{sec:Appendix-B}):
\eq{\nt 
\sigma_{xy}^{z} 
	&= - e \hbar  \int \frac{d^3k}{(2\pi)^3}  \lf( \Omega_{xy,+}^z + \Omega_{xy,-}^z \rf),
	\\ \nt %
	&= -\frac{e\hbar }{ 4\pi^2} \int_{-\infty}^\infty dk_z \int_{0}^\infty dk_\parallel k_\parallel  \lf( \Omega_{xy,+}^z  + \Omega_{xy,-}^z  \rf) 
	\\  \label{SHC-Ceq:9} 
	&= -\frac{e }{ 8\pi^2} \int_{-\infty}^\infty dk_z 
		\biggl[ \textrm{sgn}[\tilde{v}(k_z -k_0)] - \textrm{sgn}[\tilde{v}(k_z +k_0)] \biggr] 
	\\ \label{SHC-Ceq:10} 
	&=  \textrm{sgn}(m_1) \left( \frac{\hbar}{e}\right)  \frac{e^2}{h} \frac{k_0}{\pi} 
} 
with $k_\parallel=\sqrt{k_x^2+k_y^2}$.
Thus, the above value is proportional to the distance between the DPs.
This is similar to the anomalous Hall conductivity in Weyl semimetals, where the anomalous Hall conductivity depends only on the distance between Weyl points\cite{Review-WSM2018}.

\begin{figure*}[tb]
\includegraphics[width=14.8cm]{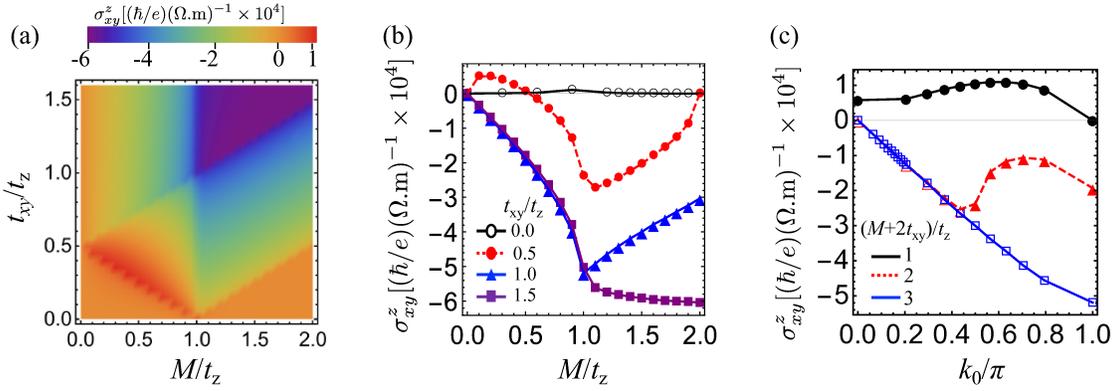}
\caption{ (Color online) 
(a) Color density plot of the SHC for $t_{xy}/t_{z}$ and $M/t_{z}$ in the lattice model.
(b) $M/t_{z}$ dependence of SHC for several $t_{xy}/t_{z}$. 
(c) $k_0$ dependence of SHC for several $(M+2t_{xy})/t_z$, which is parallel to the boundary line between TDSM and DSM phases.  
Here, $2k_0 [=2\arccos{(M-2t_{xy})/t_z}]$ is the distance between the Dirac points along the $\Gamma$-X direction.  
“Open” and “closed” symbols represent $M/t_z>1$ and $M/t_z<1$, respectively. 
Here, we used the parameters $\eta=0.89$ eV, $t_z/\eta =-3.4$, $\beta/t_z=2\gamma/t_z=0.67$, $a\sim 1$\AA, and $\mu=T=0$.
}
\label{fig:3} 
\end{figure*} 

From Eq. (\ref{SHC-Ceq:9}), the $k_z$ resolved SHC (2D SHC) $\int dk_\parallel k_\parallel (\Omega_{xy,+}^z + \Omega_{xy,-}^z) $ takes a nonzero value only between the band crossing points $k_z \in [-k_0, k_0]$ [see Fig. \ref{fig:2}(b)].
Therefore, the total SHC, which is obtained by $k_z$ integral of the nonzero $k_z$ resolved SHC, is proportional to the distance between the DPs [Fig. \ref{fig:2}(c)]. 
Later, we confirm that this feature holds even beyond the low-energy description by performing the full lattice calculation of the SHC.

\subsection{Mixing effects in low-energy description}
\label{sec:mixing}
In the previous subsection, we use the linearlized Hamiltonian in Eq.(\ref{eq:linear DH}) to evaluate the SHC. 
Here, we use the non-linearlized version in Eq.(\ref{SHC-Ceq:1}), where the mixing between DPs is taken into account.
%
The SHC is given by Eqs. (\ref{Model-eq:10})-(\ref{Model-eq:13}) as follows:
\eq{   \label{SHC-Deq:1} 
\sigma_{xy}^{z} &= e \hbar \sum_{\bm{k}}  [f(\mathcal{E}) - f(-\mathcal{E}) ] \Omega_{xy}^z,
}
where $\mathcal{E} = \sqrt{\hbar^2 v^2 (k_x^2 + k_y^2) + m^2}$ and $\Omega_{xy}^z$ denote the energy dispersion and the spin Berry curvature for the Hamiltonian $\mathcal{H}$ of Eq. (\ref{SHC-Ceq:1}), respectively.
Here, the spin Berry curvature $\Omega_{xy}^z$ is similar to that in the linearized model [Fig.\ref{fig:2}(d)], where it diverges at the DPs.
If $\bm{k}$ integral is performed in a rotation symmetric manner as before,  
the SHC at the low-temperature limit is given by
\eq{ \label{SHC-Deq:2a} 
\sigma_{xy}^{z} 
	&= - \lf(\frac{\hbar}{e} \rf) \times \frac{e^2}{h}  \int_{-\infty}^\infty \frac{dk_z}{2\pi} \mathscr{B}_{k_z}, 
	\\  \label{SHC-Deq:2b} 
\mathscr{B}_{k_z} & \equiv \int_{0}^\infty dk_\parallel k_\parallel  \hbar \Omega_{xy}^z (k_\parallel, k_z), 
}
where $\mathscr{B}_{k_z}$ is a non-dimensional value that denotes the 2D SHC of each $k_z$ or the $k_z$ resolved SHC.

First, we estimate $\mathscr{B}_{k_z}$ in the limit $m_2\to 0$ ($t_{xy}\to 0$). Then, it can be easily computed as follows (Appendix \ref{sec:Appendix-C}):
\eq{ 
\mathscr{B}_{k_z} (m_2\to0)& = \frac{1}{2} \textrm{sgn}[ - m_0 +m_1 k_z^2 ].
}
Thus, SHC becomes infinity 
\eq{   \label{SHC-Deq:3} 
\sigma_{xy}^{z} 
	& = - \textrm{sgn}(m_1) \times \infty,
}  
which can be attributed to the fact that 
$\mathscr{B}_{k_z} (m_2\to0) $ stays nonzero for large $|k_z|$ [Fig. \ref{fig:2}(e)].
This result implies that the low-energy description does not work when $m_2\to 0$.

On the contrary, $\mathscr{B}_{k_z}$ in $m_2\neq 0$ (or $t_{xy}\neq 0$) becomes an integer 
\eq{
\mathscr{B}_{k_z}
	&=  \frac{1}{2}\lf[ \textrm{sgn}(-m_0+m_1 k_z^2) -\textrm{sgn}(m_2) \rf] 
}
Here, $\mathscr{B}_{k_z}$ depends on sign of parameters.
Namely, if $m_0<0$, $m_1<0$, $m_2<0$ (e.g., parameters in Fig.  \ref{fig:1}), we have nonzero $\mathscr{B}_{k_z}$ only for $k_z \in (-k_0,k_0)$, where $2k_0 = 2\sqrt{|m_0/m_1|}$ is the distance between the DPs. 
As a result, the $k_z$ summation of $\mathscr{B}_{k_z}$, i.e., the total SHC converges [c.f., Fig. \ref{fig:2}(e)], unlike that at $m_2\to0$.
The obtained SHC is proportional to the distance between DPs.

We find that the SHC of the non-linearlized model is the same as that of the linearlized one, except when $m_2=0$. See Figs. \ref{fig:2}(c) and (f).
This is because the low-energy limit of the Hamiltonian in Eq.(\ref{Model-eq:1}) preserves the spin in the $z$-direction since the spin-orbital coupling terms $a_4({\bm k})$ and $a_5({\bm k})$ are neglected. As a result, the SHC for each $k_z$ is quantized, and thus the mixing between DPs does not affect the SHC.
%

\begin{figure*}[htb]
\includegraphics[width=13cm]{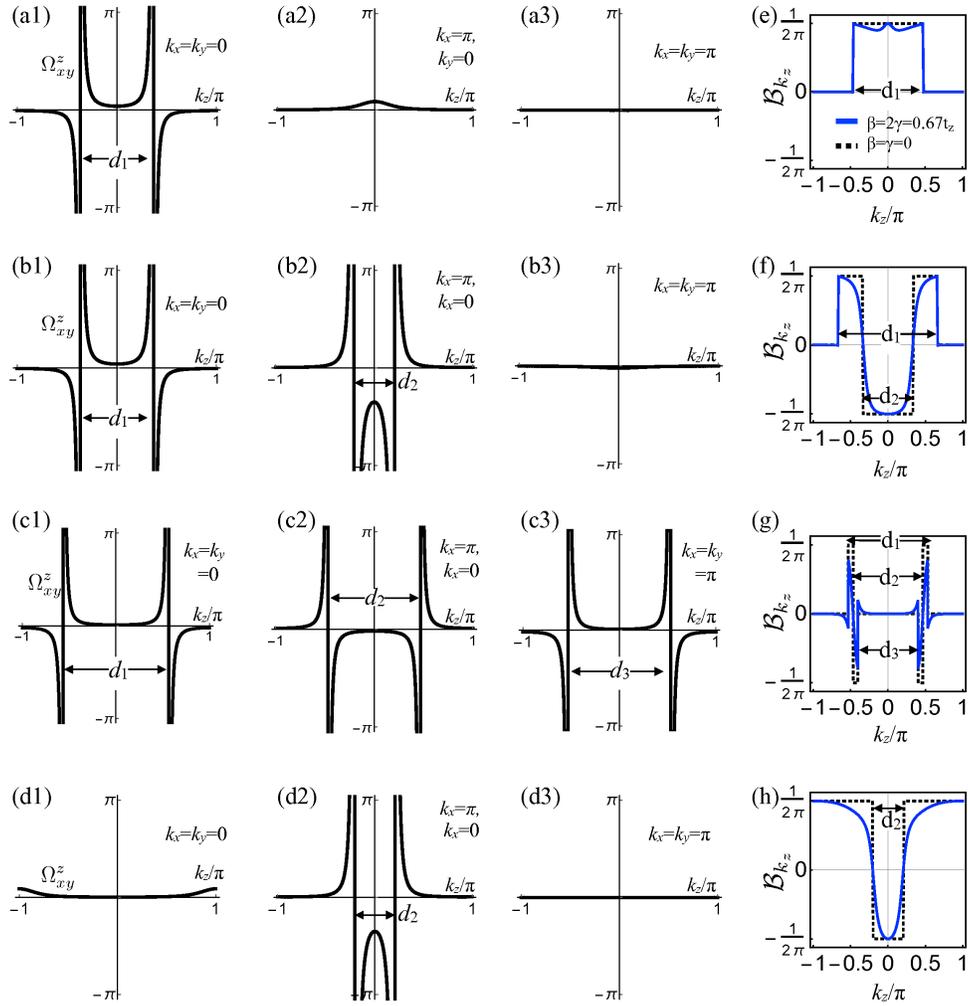}
\caption{ 
Illustration of the $k_z$ dependence of the spin Berry curvature $\Omega_{xy}^z  (\bm{k})$  
at the $\Gamma$, $X$, and $M$ points in the Brillouin zone in (a1)-(a3) TDSM phase with $M/t_z\geq1$, (b1)-(b3) TDSM with $M/t_z<1$, (c1)-(c3) DSM, and (d1)-(d3) WTI/TCI phase with $M/t_z\leq1$ at $\beta=\gamma=0$. 
Here, $d_1$, $d_2$, and $d_3$ denote the distance between the Dirac points along the $k_z$ axis at the $\Gamma$, $X$, and $M$ points, respectively. 
(e)-(h) 2D integral of the spin Berry curvature in each phase without (dotted lines) and with SOC (blue lines).
}
\label{fig:4} 
\end{figure*} 

\subsection{Full lattice description}
\label{sec:lattice}
In this section, we show the SHC in the full lattice Hamiltonian. 
Using Eq. (\ref{Model-eq:10}), we numerically compute $\sigma_{xy}^z (\mu=0)$ at $T=0$. 
The SHC depends on $t_{xy}/t_z$ and $M/t_z$, as shown in Fig. \ref{fig:3}.
The color density plot of the SHC shows several blocks, which exactly correspond to different phases in the phase diagram of Fig. \ref{fig:1}: 
in both DSM and normal insulator phases, the SHC is approximately zero.
A nonzero SHC is obtained in the TDSM phase and the WTI/TCI phase.
The phase dependence of the SHC is one of the main results in this paper.
We also confirm that the SHC shows qualitatively the same behaviors even at $T=300$ K. 

The SHC changes drastically at the line $M/t_z=1$ [see also Fig. \ref{fig:3}(b)]. 
In particular, the SHC depends in a different manner on the distance between DPs. 
Figure \ref{fig:3}(c) depicts how the SHC depends on the distance between DPs for several $(M+2t_{xy})/t_z$.
Here, $2k_0 [=\arccos{(M-2t_{xy})/t_z}]$ is the distance between the DPs along the $\Gamma$-$X$ direction,
and the “open" and “closed” symbols in Fig. \ref{fig:3}(b) denote the data for $M/t_z<1$ and $M/t_z>1$, respectively.
In contrast to the low-energy description in Secs.\ref{sec:lowenergy} and \ref{sec:mixing},
we find that the SHC is proportional to $k_0$ only when $M/t_z>1$.

This phase dependence of the SHC can be explained by the 2D integral of the spin Berry curvature 
\eq{
\mathcal{B}_{k_z} \equiv \hbar \sum_{k_x, k_y}\Omega_{xy}^z(\bm{k})  
}
in each phase.
First, we consider $\mathcal{B}_{k_z}$ at $\beta=\gamma=0$, where the spin in the $z$-direction is preserved.
Because the spin Berry curvature $\Omega_{xy}^z$ is inversely proportional to the cube of the energy dispersion $E^3$, it strongly depends on the location of the DPs of each phase. 
For example, when the DPs are located only at the $\Gamma$ point in the TDSM phases with $M/t_z>1$, 
the spin Berry curvature diverges only at the DPs along the $\Gamma$-$X$ direction [see Fig. \ref{fig:4}(a1)-(a3)].
Furthermore, its divergence is separated by $d_1$, where $d_1\equiv 2 \arccos{ (M-2t_{xy})/t_z}$ is the distance between DPs along the $\Gamma$-$X$ direction.    
Then, the non-dimensional parameter $\mathcal{B}_{k_z}$ is obtained from the distance as $\mathcal{B}_{k_z} = \frac{1}{2\pi} \textrm{sgn} (d_1 - |k_z|) $ [see Fig. \ref{fig:4}(e)].
Therefore, the SHC is proportional to $d_1$ from Eq. (\ref{Model-eq:10}) [c.f. Fig. \ref{fig:3}(b)].

On the contrary, the TDSM phase with $M/t_z<1$ hosts gapless DPs not only along the $\Gamma$-$Z$ line but also along the $X$-$R$ line. 
%
As a result, the 2D integral of the spin Berry curvature $\mathcal{B}_{k_z}$ is described by $d_1$ and $d_2$ as $\frac{1}{2\pi} [ \textrm{sgn} (d_1 - |k_z|) -2 \textrm{sgn} (d_2 - |k_z|) ]$, where $d_2=2\arccos{(M/t_z)}$ is the distance between DPs along the $X$-$A$ direction.  [Fig. \ref{fig:4}(b1)-(b3)].
The SHC depends on both $d_1$ and $d_2$, as shown in Fig. \ref{fig:4}(f). 
Thus, even in the same TDSM phase, the SHC for $M/t_z<1$
shows different behaviors than that for $M/t_z>1$. (See also  Sec.\ref{sec:IV}.)

Similarly, because the DSM phase possesses gapless DPs on the $\Gamma$-$Z$, $X$-$R$, and $M$-$A$ lines, 
the spin Berry curvature $\Omega_{xy}^z$ diverges at those points [Fig. \ref{fig:4}(c1)-(c3)].
As a result, 
$\mathcal{B}_{k_z}$ depends on $d_1$, $d_2$, and $d_3$ as follows: $\mathcal{B}_{k_z} = \frac{1}{2\pi} [ \textrm{sgn} (d_1 - |k_z|) -2 \textrm{sgn} (d_2 - |k_z|) +  \textrm{sgn} (d_3- |k_z|) ]$, where $d_3$ is the distance between the DPs along the $M$-$A$ direction. 
Because the three terms in ${\cal B}_{k_z}$ are cancelled for a finite region of $k_z$, 
the SHC is strongly suppressed in the DSM phase.
For $\beta=\gamma=0$,
we also find that the WTI/TCI phase with $M/t_z<1$ has gapless DPs along $X$-$R$ line, 
and thus the SHC in the WTI/TCI phase in $M/t_z<1$ and $M/t_z>1$ regimes are quantized to $e^2/h \times [1 - d_2/(2\pi)]$ and $e^2/h$, respectively.

In the presence of the spin-orbital coupling (SOC) with nonzero $\beta$ and $\gamma$,  the DPs along the $X$-$R$ line in $M/t_z<1$ show a narrow gap. 
Whereas the spin Berry curvature at the narrow gap  [Fig. \ref{fig:4}(b2) and (c2)] does not diverge,  it contributes to the 2D integral of the SHC, similar to that in $\beta=\gamma=0$ [Fig. \ref{fig:4}(f)-(h)].
Thus, even for nonzero $\beta$ and $\gamma$, the magnitude of SHC could be qualitatively explained by the location of the DPs and the distance between DPs. 
%
On the other hand, $\mathcal{B}_{k_z\neq0,\pi}$ is not quantized to $1/(2\pi)$ in the presence of SOC [compare Fig.   \ref{fig:4}(f) and (g)] since 
the spin in the $z$-direction is no longer conserved.
(We note that ${\cal B}_{k_z=0,\pi}$ is quantized because of the mirror reflection symmetry with respect to the $z$-axis.)


\section{Discussion}\label{sec:IV} 
First, we compare the low-energy description in Secs.\ref{sec:lowenergy}-\ref{sec:mixing} and the lattice description in Sec.\ref{sec:lattice}.
The SHC in the low-energy description is proportional to the distance between the DPs [Eq.  (\ref{SHC-Ceq:10})].
This result is consistent with that in the lattice description of the TDSM phase with $M/t_z>1$;
however, it does not agree with that with $M/t_z<1$.
This is because, in $M/t_z<1$, DPs with a tiny gap appear along the $X$-$R$ line, which provide additional contributions for 
the spin Berry curvature.
Thus, the simple low-energy description given in Sec.\ref{sec:lowenergy} is valid only in the TDSM phase with $M/t_z>1$. 

\begin{figure}[tb]
\includegraphics[width=8.5cm]{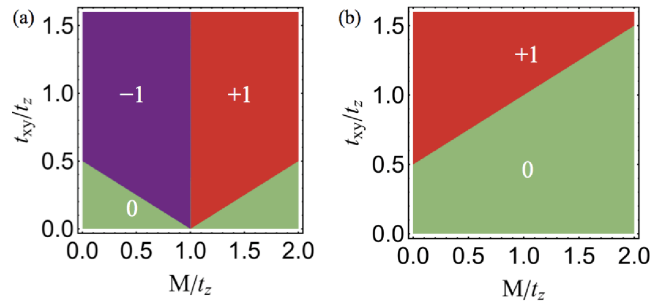}
\caption{
Phase diagram of the mirror Chern number $n_{\mathcal{M}}$ for $M/t_z$ and $t_{xy}/t_z$ at (a) $k_z=0$ and (b)$k_z=\pi$ planes.
}
\label{fig:5} 
\end{figure}

From the SHC of the lattice description, it is observed that the parameter dependence of SHC clearly changes at the line $M/t_z=1$. 
The change  can be explained by  the mirror Chern numbers for the horizontal mirror reflection symmetry. 
%
%
The mirror Chern numbers are  defined as, 
\eq{
n_{\mathcal{M}} (k_z)\equiv \frac{1}{2} \lf[ n_{+ i}(k_z)  - n_{-i}(k_z) \rf],
} 
where $k_z=0,\pi$ and $n_{\pm i}$ is the Chern number for mirror sub-sectors labeled with the mirror eigenvalues $\pm i$, which correspond to the spin-up and spin-down sectors. 
Figure \ref{fig:5} shows the mirror Chern numbers in our models, which indicates that $n_{\mathcal{M}}(k_z=0)$ changes at the line $M/t_z=1$. 
Corresponding to the nonzero mirror Chern numbers, we also have  surface states. The surface states of each phase are summarized in Appendix \ref{sec:surface states}. 

We note that the region in $M/t_z<1$ and that $M/t_z>1$ support strong and weak topological phases, respectively:
For $M/t_z<1$ ($M/t_z> 1$), 
$n_{\cal M}(k_z=0)$ and $n_{\cal M}(k_z=\pi)$ are different (the same), 
which implies a strong (weak) mirror topological  phase \cite{Chiu2013}. 
We also point out that a bulk topological index can be defined even in the presence of bulk gapless points \cite{SatoFujimoto2010}.


%

Next, we discuss the mechanism of SHC. 
Although we assume the clean limit and only consider intrinsic SHC, 
the actual system may have the extrinsic SHC due to disorders.
According to Ref. \onlinecite{Nagaosa_2010}, 
the intrinsic SHC dominates  when the longitudinal conductivity $\sigma_{xx}$ is in the range $10^4 (\Omega.\textrm{cm})^{-1} < \sigma_{xx} <10^6 (\Omega.\textrm{cm})^{-1}$. 
For Cd$_3$As$_2$, A4 and A6 samples in Ref.\onlinecite{Liang2014} satisfy this condition, which suggests that the intrinsic SHC is dominant. 
Since the band structure of Cd$_3$As$_2$ \ \cite{Wang2013a} is consistent with that in the TDSM phase with $M/t_z>1$,
our result estimates that Cd$_3$As$_2$ shows  $\sigma_{xy}^{z}\sim 10^4 (\hbar/e) \times (\Omega.\textrm{m})^{-1}$. 
The estimated value is comparable to the huge SHC in heavy metals\cite{Hoffmann2013}.

Finally, we point out that the SHC could identify phases in the phase diagram of Fig. \ref{fig:3}. 
The SHC in the TDSM and WTI/TCI phases with $M/t_z>1$ is large and negative, 
while that in DSM phase is small and negative.  
Therefore, transport measurements of the spin Hall effect can distinguish these topological phases. 

%

\section{Conclusion}
We theoretically studied the SHC in an effective model Hamiltonian of TDSM in the low-energy and the lattice descriptions. 
We found that the SHC in the linearized low-energy description is a qualitatively good approximation for the SHC in the lattice description only in the TDSM phase with $M/t_z>1$ (i.e., the DPs are located only along the $\Gamma$-$Z$ direction).   
In  addition, we found that the SHC behaves differently [for some parameters] depending on the phase in the lattice description.
It can be explained by the 2D SHC of each $ k_z $ and by the location of the DPs. 
The phase dependence of the SHC obtained in this study can be applied in determining whether a material is in the DSM or the TDSM phase by transport measurement of the spin Hall effect.

\begin{acknowledgments}
The authors thank S. Kobayashi for helpful discussion. 
This work was supported by Grant-in-Aid for Scientific Research(B) 17H02922, Early-Career Scientists 19K14658, 
the Grants-in-Aid for Scientific Research on Innovative Areas ``Topological Material Science", JSPS (Grant Nos. JP15H05855, JP15K21717, JP15H05853), and JST CREST (No: JPMJCR19T2), Japan. This project was supported in part by JSPS and ISF under Japan-Israel Cooperative Program.

\end{acknowledgments}

\ 
\newpage

\appendix

\section{Derivation of SHC}  \label{sec:Appendix-A}
In this section, we derive the SHC using Green's functions. Here, the Green's function is described as, 
\eq{ \label{appendix: eq1} 
G^\textrm{a} (\bm{k},\omega) 
	& = \frac{ \mathcal{P}_+(\bm{k}) }{\hbar\omega - i\delta - E(\bm{k})} + \frac{\mathcal{P}_-(\bm{k}) }{\hbar\omega - i\delta + E(\bm{k})}
}
where $\mathcal{P}_\pm(\bm{k}) = [1 \pm \mathcal{H}/ E(\bm{k})]/2$ denotes the projected operator, and 
$\delta >0$ is an infinitesimal value. 
The first and the second terms of the above equation imply the contribution from the dispersions of $E(\bm{k})$ and $-E(\bm{k})$, respectively. 
\begin{widetext}
Using Eq. (\ref{appendix: eq1}), we have 
\eq{ \nt 
 \sigma_{xy}^{z, \textrm{ra}} + \sigma_{xy}^{z, \textrm{aa}} 
	 &= 
	e\hbar \sum_{\bm{k},\omega} 
	\biggl\{ 
	\frac{\partial f_{\omega}}{\partial \omega} 
	\textrm{tr} \lf [ v_{\textrm{s},x}^z G^\textrm{r}_{\bm{k},\omega}  v_y G^\textrm{a}_{\bm{k},\omega}  \rf] 
	+
	 \frac{1}{2} 
	 \textrm{tr} \biggl[  v_{\textrm{s},x}^z \biggl( \frac{\partial G^\textrm{a}}{\partial \omega}  v_y G^\textrm{a} -   G^\textrm{a} v_y \frac{\partial G^\textrm{a}}{\partial \omega}  \biggr)  - \textrm{h.c.} \biggr]
	 \biggr\}
	 \\  \nt 
	& = 
	e\hbar^2 \sum_{\bm{k},\omega} 
	\biggl\{ 
	\frac{\partial f_{\omega}}{\partial (\hbar \omega)} 
	 \textrm{tr} \biggl[
		\frac{ v_{\textrm{s},x}^z \mathcal{P}_+ v_y \mathcal{P}_+ }{|h - E|^2 } + \frac{ v_{\textrm{s},x}^z \mathcal{P}_- v_y \mathcal{P}_- }{|h + E|^2 }   
	+ \frac{ v_{\textrm{s},x}^z \mathcal{P}_+ v_y \mathcal{P}_- }{ (h^* -E)(h+E)} 
		+  \frac{ v_{\textrm{s},x}^z \mathcal{P}_- v_y \mathcal{P}_+ }{ (h -E)(h^*+E)}
		\biggr]
 \\  \label{appendixB: eq2-1} 
 & \ \ \ \ \ \ \ \ \ \ \ \ \ \ \ \ \ \ \ \ \ \ \ \ +  
	 \textrm{tr} \lf[  v_{\textrm{s},x}^z \lf( \mathcal{P}_+ v_y \mathcal{P}_- - \mathcal{P}_- v_y \mathcal{P}_+ \rf)  \rf] 
 \biggl[ \frac{ f_{\omega}  E}{ [ h^2 -E^2]^2 }- \textrm{h.c.} \biggr]
	 \biggr\}, 	
}
where we used $h \equiv \hbar \omega  -i\delta $, and 
\eq{ \label{appendixB: eq2} 
\textrm{tr} \lf [ v_{\textrm{s},x}^z G^\textrm{r}_{\bm{k},\omega}  v_y G^\textrm{a}_{\bm{k},\omega}  \rf] 
	& = \textrm{tr} \biggl[
		\frac{ v_{\textrm{s},x}^z \mathcal{P}_+ v_y \mathcal{P}_+ }{|h - E|^2 } + \frac{ v_{\textrm{s},x}^z \mathcal{P}_- v_y \mathcal{P}_- }{|h + E|^2 }   
		+ \frac{ v_{\textrm{s},x}^z \mathcal{P}_+ v_y \mathcal{P}_- }{ (h^* -E)(h+E)} 
		+  \frac{ v_{\textrm{s},x}^z \mathcal{P}_- v_y \mathcal{P}_+ }{ (h -E)(h^*+E)}
		\biggr]
	\\ \label{appendix: eq3} 
 \frac{\partial G^\textrm{a} (\bm{k},\omega)  }{\partial \omega}  &= -\hbar \lf[   \frac{ \mathcal{P}_+(\bm{k}) }{[h - E(\bm{k})]^2 } + \frac{\mathcal{P}_-(\bm{k}) }{[h + E(\bm{k})]^2 } \rf] 
}
Here, 
$ \textrm{tr} \lf[  v_{\textrm{s},x}^z  \mathcal{P}_\pm v_y \mathcal{P}_\pm  \rf] $ and $ \textrm{tr} \lf[  v_{\textrm{s},x}^z  \mathcal{P}_\pm v_y \mathcal{P}_\mp  \rf] $ in Eq. (\ref{appendixB: eq2-1}) are estimated from 
\eq{  \label{appendixB: eq3} 
\textrm{tr} \lf [  v_{\textrm{s},x}^z v_y \rf] &= 0
	\\  \label{appendixB: eq4}
\textrm{tr} \lf [ v_{\textrm{s},x}^z  \mathcal{H} v_y  \mathcal{H} \rf] &= 0
	\\ \label{appendixB: eq5}
\textrm{tr} \lf [\mathcal{H} (v_{\textrm{s},x}^z v_y + v_y v_{\textrm{s},x}^z  ) \rf] &= 0
}
and 
\eq{ \label{appendixB: eq7} 
&\textrm{tr} \lf [  v_{\textrm{s},x}^z \mathcal{P}_\pm v_y \mathcal{P}_\pm \rf] 
	 =  \frac{1}{4} \textrm{tr}  \lf[  v_{\textrm{s},x}^z v_y  +  v_{\textrm{s},x}^z  \mathcal{H} v_y  \mathcal{H}  \pm \frac{1}{E} \mathcal{H} (v_{\textrm{s},x}^z v_y + v_y v_{\textrm{s},x}^z  ) \rf]
= 0,
	\\ \label{appendixB: eq8}
&\textrm{tr} \lf [  v_{\textrm{s},x}^z \mathcal{P}_\pm v_y \mathcal{P}_\mp \rf] 
	 =  \frac{1}{4} \textrm{tr}  \lf[  v_{\textrm{s},x}^z v_y  -  v_{\textrm{s},x}^z  \mathcal{H} v_y  \mathcal{H}  \pm \frac{1}{E} \mathcal{H} (v_{\textrm{s},x}^z v_y - v_y v_{\textrm{s},x}^z  ) \rf] 
	=  \pm \frac{1}{4E} \textrm{tr} \lf[ \mathcal{H} (v_{\textrm{s},x}^z v_y - v_y v_{\textrm{s},x}^z  ) \rf].
}
Here, Eq. (\ref{appendixB: eq8}) is also represented by the spin Berry curvature $\Omega_{xy}^z (\bm{k})$:
\eq{ \label{appendix: eq11} 
 \textrm{tr} \lf[  v_{\textrm{s},x}^z  \mathcal{P}_\pm v_y \mathcal{P}_\mp  \rf] 
 	& =  \pm 2i   \Omega_{xy}^z (\bm{k}) E^2 (\bm{k}).
}
Here, the spin Berry curvature is defined by 
\eq{\nt 
\Omega_{xy}^z &\equiv  \sum_{m\neq n, E_n> E_{m}} i
 \frac{ \langle n, \bm{k} | v_{\textrm{s},x}^z  | m, \bm{k}  \rangle \langle m, \bm{k} | v_y  | n, \bm{k}  \rangle - (x \leftrightarrow y) }{ [ E_n(\bm{k}) - E_m(\bm{k})]^2  }
	\\ \label{appendix: eq10} 
	& =   \frac{ \eta^2 }{2\hbar E^3} 
		\biggl[
		\cos{k_x} \cos{k_y} (M - t_{z} \cos{k_z}) 
		- t_{xy} (\cos{k_x} + \cos{k_y})     \biggr],
}
where $ | m, \bm{k}  \rangle$ denotes the wave function for $E_m$ of the $m$-th band.
The spin Berry curvature is given by $ \textrm{tr} \lf[  v_{\textrm{s},x}^z  \mathcal{P}_\pm v_y \mathcal{P}_\mp  \rf] $.
As a result, we have,
\eq{ 
& \sigma_{xy}^{z, \textrm{ra}} + \sigma_{xy}^{z, \textrm{aa}} 
	 = 
	e\hbar^2 \sum_{\bm{k},\omega}  \lf[ 2i E^2 \Omega_{xy}^z (\bm{k})  \rf]
	\biggl\{ \frac{\partial f_{\omega}}{\partial (\hbar \omega)}   \lf [ 
		 \frac{1 }{ (h^* -E)(h+E)} 
		-\textrm{h.c.}
		\rf]
		+ 2 
	\biggl[ \frac{ f_{\omega}  E}{ [ h^2 -E^2]^2 }- \textrm{h.c.} \biggr]
	 \biggr\}   .
}
Here, the $\omega$ integral in the above equation is obtained as follows: 
\eq{ \nt 
& \sum_{\omega}  
	\biggl\{
	\frac{\partial f_{\omega}}{\partial (\hbar \omega)}   \lf [ 
		 \frac{1 }{ (h^* -E)(h+E)} 
		- \textrm{h.c.}
		\rf]
		+2 
	\biggl[ \frac{ f_{\omega}  E}{ [(\hbar\omega - i\delta )^2 -E^2]^2 }- \textrm{h.c.} \biggr]
	\biggr\}
	\\
\nt 
	 = & \frac{1}{2\pi \hbar } \int_{-\infty}^\infty d\hbar\omega  
	\biggl[ \frac{\partial f_{\omega}}{\partial (\hbar \omega)}   \lf [ 
		 \frac{1 }{ (\hbar\omega +i\delta -E)(\hbar\omega - i\delta +E)} 
		- \textrm{h.c.}
		\rf]
		+ 2  
	\biggl[ 
	\frac{ f_{\omega}  E}{ [ (\hbar\omega - i\delta ) -E ]^2 [ (\hbar\omega - i\delta ) + E ]^2 }
	- \textrm{h.c.}
	 \biggr]   
	 \biggr\}
	\\ \nt 
	 =& \frac{1}{2\pi \hbar } \oint_{\textrm{upper half circle}} dz   
	\biggl\{ f'_z  \lf [ 
		 \frac{1 }{ [z - ( E-i\delta)] [z - (-E+i\delta)]} 
		- \frac{1 }{ [z - ( E+i\delta)] [z - (-E-i\delta)]} 
		\rf]
	\\ \nt 
	& \  \ \ \ \ \ \ \ \ \ \ \ \ \  \ \ \ \ \  \ \ \ \ \ 
	+ 2 
	\biggl[ 
	\frac{ f_z  E}{ [ z - (E+ i\delta) ]^2 [ z - (-E+ i\delta)   ]^2 }
	-
	\frac{ f_z  E}{ [ z - (E- i\delta) ]^2 [ z - (-E- i\delta)   ]^2 }
	 \biggr] 
	 \biggr\}  
	\\  \label{appendix: eq5} 
	=& \frac{i}{2\hbar}
		\lf[ 
		\frac{f'_z(-E+i\delta)}{-E+i\delta} - \frac{f'_z(E+i\delta)}{E+i\delta} 
		\rf] 
		 +2\times  \frac{i}{4\hbar }  \lf[ \frac{f'(E) + f'(-E)}{E}  - \frac{f(E) - f(-E)}{E^2}\rf]
	\\
	  \label{appendix: eq5-2} 
	 =&_{\delta\to0} 
	  - \frac{i}{2\hbar}
		\frac{f'_z(E)+f'_z(-E)}{E}
		 +2\times  \frac{i}{4\hbar }  \lf[ \frac{f'(E) + f'(-E)}{E}  - \frac{f(E) - f(-E)}{E^2}\rf]
	\\
	  \label{appendix: eq5-3} 
	=& -\frac{i}{2\hbar } \frac{f(E) - f(-E)}{E^2} 	
}
As a result, we obtain
\eq{   \label{appendix-A:eq6}
\sigma_{xy}^{z, \textrm{ra}} + \sigma_{xy}^{z, \textrm{aa}} 
	& = 
	e\hbar \sum_{\bm{k}} [f(E) - f(-E)] \Omega_{xy}^z (\bm{k}).
}
Here, the SHC associates the Fermi sea term, according to Strede formal\cite{Tanaka2008,Streda1982}. 

\section{Calculation of SHC in the low-energy description}
\label{sec:Appendix-B}
Here, we show the SHC in the low-energy description. The SHC is obtained from the spin Berry curvature in the effective model Hamiltonian $\mathcal{H}(\bm{k})$ 
by replacing $\sin{k_i} \to k_i a +o(k_i^3)$, $\cos{k_i} \to 1- \frac{1}{2}k_i^2 a^2+o(k_i^3)$, and $\eta \to \hbar v$ in Eq. (\ref{Model-eq:2})-(\ref{Model-eq:4}) as follows:
\eq{ \nt
\Omega_{xy}^{z}  (k_x, k_y,k_z)
	& = \frac{\hbar v^2  }{2}  \frac{ m(\bm{k})} {E^3}
}  
where we used $\eta a = \hbar v$ and $m (\bm{k})= -m_0 + m_1 k_z^2 + m_2 (k_x^2 +k_y^2)$.
Here, the spin Berry curvature $\Omega_{xy}^{z}  (k_x, k_y,k_z)$ can be estimated near the Dirac points $\bm{k}_\pm =(0,0, \pm k_0)$. 
Then, the spin Berry curvature linearized $\bm{k}$ near the DPs $\Omega_{xy,\pm}^{z}$ is given by 
the energy dispersion $E_\pm (\bm{k} \sim \bm{k}_\pm) = \hbar \sqrt{v^2 (k_x^2 + k_y^2) + \tilde{v}^2 (k_z \mp k_0)^2  }$ as follows: 
\eq{
\Omega_{xy,\pm}^{z}  (\bm{k} \sim \bm{k}_\pm)
	& =  \pm \frac{\hbar v^2}{2}  \frac{   \hbar \tilde{v} (k_z \mp k_0) } {E_\pm^3}
	\\
	& =  \pm \frac{\hbar v^2}{2}  \frac{  \hbar \tilde{v} (k_z \mp k_0) } { [\hbar^2 v^2 (k_x^2 + k_y^2) + \hbar^2 \tilde{v}^2 (k_z \mp k_0)^2]^{3/2}}
}  
As a result, we obtain 
\eq{\nt 
\sigma_{xy}^{z} 
	&= -e\hbar \sum_{\pm } \int \frac{dk^3}{(2\pi)^3} \Omega_{xy,\pm}^z (k_x,k_y, k_z)
	\\ \nt
	&  = - \frac{e\hbar}{ 4\pi^2}\sum_{\pm } \int_{-\infty}^\infty dk_z \int_{0}^\infty dk_\parallel k_\parallel  \Omega_{xy,\pm}^z (k_\parallel, k_z)
	\\ \nt
	&= -\frac{e\hbar}{ 4\pi^2}  \sum_{\pm }\int_{-\infty}^\infty dk_z \int_{0}^\infty dk_\parallel k_\parallel 
		\lf[  \pm \frac{\hbar v^2}{2}  \frac{  \hbar \tilde{v} (k_z \mp k_0) } { [\hbar^2 v^2 k_\parallel^2  + \hbar^2 \tilde{v}^2 (k_z \mp k_0)^2]^{3/2}} \rf]
	\\ \nt 
	& = -\frac{e}{ 8\pi^2}  \sum_{\pm }\int_{-\infty}^\infty dk_z \int_{0}^\infty d \epsilon  
		\lf[  \pm   \frac{  \hbar \tilde{v} (k_z \mp k_0) \epsilon   } { [ \epsilon^2   + \hbar^2 \tilde{v}^2 (k_z \mp k_0)^2]^{3/2}} \rf]
	\\ 
	&= -\frac{e }{ 8\pi^2} \int_{-\infty}^\infty dk_z 
		\biggl[ \textrm{sgn}[ \tilde{v}(k_z -k_0)] - \textrm{sgn}[\tilde{v}(k_z +k_0)] \biggr] 
	\\ 
	&= \textrm{sgn}(\tilde{v})\frac{ek_0 }{ 2\pi^2}
		\\ 
	&= \textrm{sgn}(m_1)  \lf( \frac{\hbar}{e} \rf) \frac{e^2}{2\pi \hbar} \frac{k_0 }{ \pi}, 
}  
where we used $\tilde{v}= \textrm{sgn}(m_1) 2 \sqrt{m_0m_1}/\hbar$.

\section{Calculation of SHC in the low-energy description with the mixing effect}
\label{sec:Appendix-C}
In this section, we calculate the SHC in the quadratic Hamiltonian. 
The SHC is given by Eq. (\ref{SHC-Ceq:1}) as  
\eq{
\sigma_{xy}^{z} 
	&= -\frac{e\hbar}{ 4\pi^2 }  \int_{-\infty}^\infty dk_z \int_{0}^\infty dk_\parallel k_\parallel  \Omega_{xy}^z (k_\parallel, k_z)
	\\
	&= - \lf(\frac{\hbar}{e} \rf) \times \frac{e^2}{h}  \int_{-\infty}^\infty \frac{dk_z}{2\pi} \mathcal{B}_{k_z} 
}
with
\eq{
\mathcal{B}_{k_z} & \equiv \int_{0}^\infty dk_\parallel k_\parallel  \hbar \Omega_{xy}^z (k_\parallel, k_z), 
    \\
\Omega_{xy}^{z}  (\bm{k})
	& = \frac{\hbar v^2  }{2}  \frac{ -m_0 + m_1 k_z^2 - m_2 (k_x^2+k_y^2) } {E^3}
	 \\
E (\bm{k} ) &= \sqrt{\hbar^2 v^2 (k_x^2 + k_y^2) + m^2 (\bm{k})}     \\
	m(\bm{k}) & = -m_0 + m_1 k_z^2 + m_2 (k_x^2+k_y^2)
}
If $m_2=0$ (or $t_{xy}=0$), because $m(\bm{k})$ is independent of $k_x$ and $k_y$, $\sigma_{xy}^{z} $ is easily calculated as follows: 
\eq{ 
\mathcal{B}_{k_z} & =\frac{1}{2}\textrm{sgn} [-m_0 + m_1 k_z^2] 
\\
\sigma_{xy}^{z} 
	&= - \lf(\frac{\hbar}{e} \rf) \times \frac{e^2}{h}  \int_{-\infty}^\infty \frac{dk_z}{2\pi} \frac{1}{2}\textrm{sgn} [-m_0 + m_1 k_z^2]   
	\to - \textrm{sgn}(m_1) \times  \infty.
}

\section{Surface states}\label{sec:surface states}
Because the nonzero mirror Chern number ensures surface states, we consider the surface states of each phase [$k_x$ surface at $k_z=0$ and $k_z=\pi$].
As a result, the surface states in WTI/TCI of $M/t_z<1$ and $M/t_z>1$ are different, as shown in Fig.\ref{fig:A1}.
It is further found that surface states exist in WTI/TCI and TDSM phases. 
In the WTI/TCI phase $M/t_z<1$ ($M/t_z>1$), the band crossing points of the surface states at $k_z=0$ and $k_z=\pi$ are present at different locations (at the same location $k_y=0$).
Furthermore, the band crossing points are located at symmetric positions,
which could have been caused by mirror-reflection symmetry. 

\begin{figure}[tb]
\includegraphics[width=17.8cm]{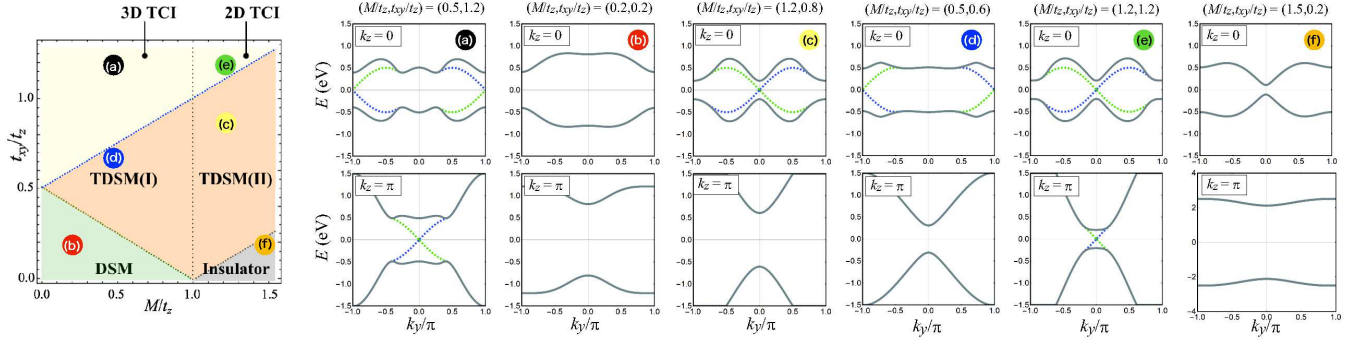}
\caption{
(100) surface states at (upper panels) $k_z=0$ and (lower panels) $k_z=\pi$ in WTI/TCI phase of (a) $M/t_z<1$ and (b) $M/t_z>1$, in TDSM phase of (c) $M/t_z<1$ and (d) $M/t_z>1$, and in (e) DSM phase.
Blue and green dotted lines represent the surface states of up-spin and down-spin sectors, respectively. The left panel shows the phase diagram.
}
\label{fig:A1} 
\end{figure} 
\end{widetext}



\end{document}